\documentclass[a4paper,11pt]{article}
\begin{document}
\def\baselinestretch{1.5}
{\def\title{The approach to the analysis of the dynamic of
non-equilibrium open systems}
\def\author{V.M. Somsikov}
\def\correspondence{Prof. V.M. Somsikov}
\def\correspondenceshort{Prof. Somsikov}
\def\date{\today}
\def\addra{Laboratory of Physics of the geogeliocosmic relation}
\def\addrb{Institute of Ionosphere}
\def\addrc{Kamenskoe Plato}
\def\addrd{Almaty, 480020, Kazakhstan}
\def\tel{~~  ~~~+8-3272-548074~~~~~~~~~~~~~~~~~~~~~}
\def\fax{~~  ~~~+8-3272-658085~~~~~~~~~~~~~~~~~~~~~}
\def\email{~~  ~~ nes@kaznet.kz ~~~~~~~~~~}}
\small
%\column\hsize\textwidth\columnwidth\hsize\csname
%@columnfalse\endcsname
\title {The approach to the analysis of the
dynamic of non-equilibrium open systems and irreversibility}
\author{Vyacheslav Somsikov }
\date{\it{Laboratory of Physics of the geogeliocosmic relation,\\
Institute of Ionosphere, Almaty, 480020, Kazakhstan}} \maketitle
\begin{abstract}
The approach to the analysis of the dynamic of non-equilibrium
open systems within the framework of the laws of classical
mechanics on the example of a hard-disks is offered. This approach
was based on Hamilton and Liouville generalized equations which
was deduced for the subsystems of the nonequilibrium system. With
the help of generalized Liouville equation it was obtained that
two types of dynamics are possible: reversible and irreversible.
The connection between the dynamical parameter -generalized field
of forces, and entropy is established. The estimation of
characteristic time of establishment of equilibrium in the
thermodynamic limit is realized. It is shown how from the
condition of irreversibility of a hard-disk system, the condition
of irreversibility for the rarefied system of potentially
interacting particles follows. The explanation of the mechanism of
irreversibility is submitted.
\end{abstract}
\bigskip

\date{\it{Introduction.}} \maketitle
In connection with a modern physics, the Newton's equation is a
background of dynamical picture of the world. It is caused by the
fact that all four known fundamental interactions of elementary
particles are potential. Potentiality of forces causes
reversibility of the Newton equation in time. Hence, dynamics of
all natural systems consisting of elementary particles should be
reversible. But irreversibility is a basis of evolution. In
fundamental physics irreversibility determines the contents of the
second law of thermodynamics. According to this law there is
function S named entropy, which can grow for isolated systems
only, achieving a maximum in an equilibrium state. Thus, the
fundamental physics includes areas contradicting each other:
reversible classical mechanics and irreversible thermodynamics.

So, the rigorous substantiation of the second law of
thermodynamics within the framework of the classical mechanics
laws is one of the primary tasks of modern physics. This problem
keenly reveals difficulties and limits of capability resources of
classical methods of physics. Despite of numerous attempts, this
problem does not have satisfactory solution up to date. Thus, the
question how one can describe irreversible processes on the basis
of the reversible equations of the classical mechanics, are
discussed and studied [1, 5, 8, 13,14]. The solution of this
question is necessary for development of the theory of
irreversible processes on dynamic basis. In order to find an
answer for this question, we studied the evolution of the
non-equilibrium colliding hard-disk system [10-12].

Here we are going to show, how one can study of the mechanism of
establishment of the equilibrium state within the framework of the
classical mechanics laws if one replaces the canonical Hamilton
and Liouville equations with the generalized equations, applied
for nonholonomic and open systems.

The main idea of our approach to analyzes of non-equilibrium
system is splitting of conservative nonequilibrium system into
interacting subsystems, research of dynamics of one of them with
help of the generalized Liouville equation, is not brought to the
assumption of potentiality of their forces of interaction.

The work is constructed as follows. Using D'Alambert and the
motion equations for disks, we obtain generalized Lagrange,
Hamilton and Liouville equations for disk subsystem, which is
selected from the whole disk system. Then the analysis of two
types of evolution of nonequilibrium system - reversible and
irreversible, was made. The existence of these types of evolution
follows from the condition of compatibility of the generalized
Liouville equation for subsystem and the canonical Liouville
equation, applicable for whole system. Estimation of
characteristic parameter of establishment of the equilibrium state
was realized. Then, it is shown how the condition of
irreversibility for the potentially interacting and rarefied disks
system follows from the condition of irreversibility for a
hard-disk system. The formula that determines the connection
between increment entropy for full non-equilibrium system and the
generalized field of forces of subsystems was obtained. In
summary, the analysis of the results is submitted.

 \date{\it{Mathematical apparatus.}} \maketitle
We used an approach of pair interaction of disks. Their motion
equations are deduced with help of the matrix of pair collisions.
As complex plane this matrix is given [10]:

$S_{kj}=\left(\begin{array}{cc} a & -i b
\\ -i b & a\end{array} \right)$,
where $a=d_{kj}\exp(i\vartheta_{kj})$; $b=\beta
\exp(i\vartheta_{kj})$; $d_{kj}=cos\vartheta_{kj}$;
$\beta=sin\vartheta_{kj}$; $i$- an imaginary unit; $k$- and $j$-
numbers of colliding disks; $d_{kj}$- the impact parameter (IP),
determined by distance between centers of colliding disks in the
Cartesian plane system of coordinates with axes of $x$ and $y$, in
which the $k$-disk swoops on the lying the $j$- disk along the $x$

- axis. The scattering angle $\vartheta_{kj}$ varies from $0$ to
$\pi$. In consequence of collision the transformation of disks
velocities can be presented in such form: $V_{kj}^{+}=S_{kj}
V_{kj}^{-}$ (a), where $V_{kj}^{-}$ and $V_{kj}^{+}$ - are
bivectors of velocities of $k$ and $j$ - disks before $(-)$, and
after $(+)$ collisions, correspondingly; $V_{kj}$=$\{V_k,V_j\}$,
${V}_j=V_{jx}+iV_{jy}$ - are complex velocities of the incident
disk and the disk - target with corresponding components to the
$x$- and $y$- axes. The collisions are considered to be central,
and friction is neglected. Masses and diameters of disks-$"d"$ are
accepted to be equal to $1$. Boundary conditions are given as
either periodical or in form of hard walls. From (a) we can obtain
equations for the change of velocities of colliding disks [10-12]:
\begin{equation}
{\left(\begin{array}{c} {\delta V_k}\\ {\delta V_j}
\end{array} \right )
=\varphi_{kj} \left( \begin{array}{c} \Delta_{kj}^{-} \\
-\Delta_{kj}^{-}\end{array} \right )}.\label{eqn1}
\end{equation}
Here, $\Delta_{kj}=V_k-V_j$ - are relative velocities, $\delta
V_k=V_k^{+}-V_k^{-}$, and $\delta V_j=V_j^{+}-V_j^{-}$ - are
changes of disks velocities in consequence of collisions,
$\varphi_{kj}=i\beta e^{i\vartheta_{kj}}$. \\That is, Eq. (1) can
be presented in the differential form of Somsikov, V.M. [2001]:
\begin{equation}
\dot{V}_k=\Phi_{kj}\delta (\psi_{kj}(t))\Delta_{kj}\label{eqn2}
\end{equation}
where $\psi_{kj}=[1-|l_{kj}|]/|\Delta_{kj}|$;
$\delta(\psi_{kj})$-delta function; $l_{kj}(t)=z_{kj}^0+\int
\limits_{0}^{t}\Delta_{kj}{dt}$ - are distances between centers of
colliding disks; $z_{kj}^0=z_k^0-z_j^0$, $z_k^0$ and $ z_j^0$ -
are initial values of disks coordinates;
$\Phi_{kj}=i(l_{kj}\Delta_{kj})/(|l_{kj}||\Delta_{kj}|)$.

    The Eq. (2) determines transformation of velocity of $k$ -disk
when it collides with $j$-disk. A right side of {Eq}.(2)
represents force depending from $\Delta_{kj}$. Therefore the
equation (2) is not a Newtonian one [4].

    Let's take a disk system, which consist of $N$ disks, when
$N\rightarrow\infty$ and $L^{2}\rightarrow\infty$, where
$N/L^{2}=finit$; and $L^{2}$ - is area occupied by disks. Then
divide this system on $R$ subsystems, so that in each subsystem
will be $T>>1$ disks. Therefore, $N=RT$. Let energy of the system
is equal $E=const$. It is equal to the sum of internal energies of
all subsystems and interaction energies between subsystems. Let us
select one of them, which we call $p$-subsystem. Let us examine
$\delta{W}_{a}^{p}$ - the virtual work of active forces made above
$p$-subsystem. In the general case, this work can be presented us
follows: $\delta{W}_{a}^{p}=\sum\limits_{k=1}^{T}
\sum\limits_{s=1}^{N-T}F_{ks}^p\delta{r_k}= \sum\limits_{k=1}^{T}
F_{k}^p\delta{r_k}$, where $k=1,2,3...T$ -disks number of the $p$
- subsystem, $s=1,2,3...N-T$ - external disk number to the $p$
-subsystem of the disk, which interaction with $k$ -disk of
$p$-subsystem, $F_{ks}^p$ - the interaction force of  $k$ - and
$s$ - disks, $\delta{r_k}$ - virtual displacement of the $k$
-disk, $F_{k}^p=\sum\limits_{s=1}^{N-T}F_{ks}^p$. Here we use that
the virtual work of the interaction force of internal disks of a
$p$-subsystem is equal to zero.

    In case of the pair interaction approaching, the virtual work of
external forces gets the form: $\delta{W}_{a}^{p}=
\sum\limits_{k=1}^{T}
F_{k}^p\delta{r_k}=\sum\limits_{k=1}^{T}F_{ks}^p\delta{r_k}$,
because in this case $F_{k}^p = F_{ks}^p$. The inertial force can
be presented so:
$\delta{W}_{in}^p=\sum\limits_{k=1}^T\dot{V}_k\delta{r_k}$. The
sum of the active and the inertial forces is called the effective
force. The principle of D'Alambert asserts that the work of
effective forces is always equal to zero [2], i.e.
\begin{equation}
{\delta\overline{W}_q^p=\delta{W}_{in}^p-\delta{W}_a^p}.\label{eqn3}
\end{equation}
The feature above virtual work means, that in generally it is not
reduced to a complete differential.

    From the motion equations for a hard disks it
follows [11]:
$\sum\limits_{p=1}^{R}(\sum\limits_{k=1}^{T}F_{ks}^p)\delta{r_k}=0$.
Therefore the total of the active, and the total inertial works
for all subsystems at any moment of time is equal to zero, i.e.
$\sum\limits_{p=1}^R{\delta{W}_{in}^p}=
\sum\limits_{p=1}^R{\delta{W}_a^p}=0$. This equaling can take
place in two cases: when the sum of nonzero members is equal to
zero, and when each member of the sum is equal to zero. It is
obviously that the second case, appropriate to an equilibrium
state, takes place when $T\rightarrow\infty$. For this case, with
the help of the motion equations for a hard disks, it is possible
to record:
\begin{equation}
\sum\limits_{k=1}^T\dot{V_k}=\sum\limits_{k=1}^T\Phi_{ks}\delta
(\psi_{ks}(t))\Delta_{ks}(t)= \sum\limits_{k=1}^T{F_{ks}^p}=0,
\label{eqn4}
\end{equation}

    Equality to zero of the right-hand side of the equation (4) means,
that the selected $p$-subsystem is in a stationary state.

    To obtain the general Lagrange equation for $p$-subsystem, let's
transform D'Alambert equation (3) by multiplied it by $dt$, and
integrated it over an interval from $t=t_1$ to $t=t_2$. In the
general case we have:
$\int\limits_{t_1}^{t_2}{\delta{\bar{W_q^p}}}dt$=
$\int\limits_{t_1}^{t_2}{\sum\limits_{k=1}^T[{\frac{d}{dt}}V_k-
\sum\limits_{j\neq k}^T{F_{kj}^p}-F_{k}^p]\delta{r_k}dt}$ =

$\delta\int\limits_{t_1}^{t_2}
{\frac{1}{2}}\sum\limits_{k=1}^T{V_k^2}dt-
\int\limits_{t_1}^{t_2}[\sum\limits_{k=1}^T(F_k^p+
\sum\limits_{j\neq k}^T F_{jk}^p)\delta{r_k}]dt-
[\sum\limits_{k=1}^T{V_k}{\delta{r_k}}]_{t_1}^{t_2}$ (b)
    In equation (b) the member $\sum\limits_{j\neq k}^T{F_{kj}^p}$
determines the force of interaction a internal disks of the
$p$-subsystem. Then $k$ and $j$ -colliding disks from the
$p$-subsystem. The member $F_k^p$ is the force on the
$p$-subsystem. If it is demanded that on the ends an interval
$[t_1, t_2]$ the virtual displacements are zero, the last member
in (b) will be equal to zero.

    Let's neglected nonequilibrium inside a subsystem (this is a
typical assumption for local equilibrium). Then for internal
forces of interaction of subsystem disks, we can set in the
conformity such a function dependent on coordinates,
$U(r_1,r_2,...r_T)$, for which the following condition is
satisfied:
$\int\limits_{t_1}^{t_2}[\sum\limits_{k=1}^T{\sum\limits_{j\neq
k}^T{F_{kj}^p}}\delta{r_k}]dt=
-\delta\int\limits_{t_1}^{t_2}U(r_1,r_2,...r_T)dt$. Here -
$r_1,r_2...r_T$ - coordinates of disks of the $p$-subsystem disks.
In general case it is impossible, to present acting on
$p$-subsystem active forces, as a gradient of a force function
[2]. In this case the equation (b) can be written as:
\begin{equation}
\int\limits_{t_1}^{t_2}\delta{\bar{W_q^p}dt}=
\int\limits_{t_1}^{t_2}
[\sum\limits_{k=1}^T({\frac{d}{dt}}{\frac{\partial{L_p}}
{\partial{V_k}}}-{\frac{\partial{L_p}}
{\partial{r_k}}}-{F_{k}^p}){\delta{r_k}}]{dt}=0 \label{eqn5}
\end{equation}
In eqn. (5) we denote
$L_p=\sum\limits_{k=1}^T{\frac{V_k^2}{2}}+U(r_1,r_2,..r_T)$.
Therefore, if the interaction of disks will be potential, the
$L_p$ will include also internal potential energy of the
$p$-subsystem - $U(r_1,r_2,...r_T)$. Because for any variations
integral in equation (5) will be equal to zero, the next
expression is carried out:
\begin{equation}
\sum\limits_{k=1}^T(\frac{d}{dt}\frac{\partial{L_p}}
{\partial{V_k}}-\frac{\partial{L_p}}{\partial{r_k}})=\sum\limits_{k=1}^T{F_k^p}=F_p
\label{eqn6}
\end{equation}
We denote $\sum\limits_{k=1}^T{F_k^p}=F_P$.

    The equation (6) is a generalized equation of Lagrange for a
$p$-subsystem. So, $F_p$, is the polygenic force acting on the
$p$-subsystem which dependent from its dynamic. When the
$F_{p}=0$, the equation (6) transforms to a canonical equation of
Lagrange for an equilibrium, conservative system.

    Let us derive Hamilton's equations for $p$-subsystem. The
differential for $L_p$ can be written as:

$dL_p=\sum\limits_{k=1}^T(\frac
{\partial{L_p}}{\partial{r_k}}dr_k+\frac{\partial{L_p}}
{\partial{V_k}} d{V_k}) +\frac{\partial{L_p}}{\partial{t}}dt$,
where $\frac{\partial{L_p}}{\partial{V_k}}=p_k$ - is disks
momentum. With the help of Lagrange transformation, it is possible
to get: $d[\sum\limits_{k=1}^Tp_k{V_k}-L_p]=
\sum\limits_{k=1}^T(-\frac{\partial{L_p}}{\partial{r_k}}dr_k+
{V_k}dp_k)-\frac{\partial{L_p}}{\partial{t}}dt$. Because
$\frac{\partial{H_p}}{\partial{t}}=-\frac{\partial{L_p}}{\partial{t}}$,
where $H_p=\sum\limits_{k=1}^Tp_k{V_k}-L_p$, we will have from
(6):
\begin{equation}
{\frac{\partial{H_p}}{\partial{r_k}}=-\dot{p_k}+F_{k}^p}
\label{eqn7}
\end{equation}
\begin{equation}
{\frac{\partial{H_p}}{\partial{p_k}}={V_k}}\label{eqn8}
\end{equation}
These are the general Hamilton equations for the selected
$p$-subsystem. The external forces, which acted on $p$-subsystem,
presented in a right-hand side an equation (7).

    Using equations (7,8), we can find the Liouville equation for
$p$-subsystem. For this purpose, let's to take a generalized
current vector - $J_p=(\dot{r_k},{\dot{p_k}})$ of the
$p$-subsystem in a phase space [13]. From equations (7,8), we
find:
\begin{equation}
{divJ_p=\sum\limits_{k=1}^T({\frac{\partial}{\partial{r_k}}}{V_k}+
\frac{\partial}{\partial{p_k}}{\dot{p_k}})=
\sum\limits_{k=1}^T{\frac{\partial}{\partial{p_k}}{F_{k}^p}}}
\label{eqn9}
\end{equation}

    The differential form of the particle number conservation law in
the subsystem is a continuity equation:
$\frac{\partial{f_p}}{\partial{t}}+{div(J_pf_p)}=0$, where
$f_p=f_p(r_k,p_k,t)$- the normalized distribution function of
disks in the $p$-subsystem. With the help of the continuity
equation and equation (9) for a divergence of a generalized
current vector in a phase space, we can get:
$\frac{df_p}{dt}=\frac{\partial{f_p}}{\partial{t}}+\sum\limits_{k=1}^T({V_k}
\frac{\partial{f_p}}{\partial{r_k}}+\dot{p_k}\frac{\partial{f_p}}
{\partial{p_k}})=\frac{\partial{f_p}}{\partial{t}}+div(J_pf_p)-f_pdivJ_p=
-f_p\sum\limits_{k=1}^T\frac{\partial}{\partial{p_k}}F_{k}^p$. So,
we have:
\begin{equation}
{\frac{df_p}{dt}=-f_p\sum\limits_{k=1}^T
\frac{\partial}{\partial{p_k}}F_{k}^p} \label{eqn10}
\end{equation}

Equation (10) is a Liouville equation for $p$-subsystem. It has a
formal solution:
\\${f_p=const\cdot{\exp{[-\int\limits_{0}^{t}{(\sum\limits_{k=1}^T
\frac{\partial}{\partial{p_k}}F_{k}^p)}{dt}]}}}$.

The equation (10) is obtained from the common reasons. It is
suitable for any interaction forces of subsystems. Thus, the
equation (10) is applicable to analyze any nonequilibrium open
systems. In particularly, it can be used for explanation of
irreversibility. The right side of (10), $\sum\limits_{k=1}^T
\frac{\partial}{\partial{p_k}}F_{k}^p$, is the integral of
collisions. This integral can be obtained from the motion
equations of the systems element. For example, for a hard disks
system it can be found with the help of the Eq. (2).

    The stationary nonequilibrium state of the system can be supported
by external constant stream of energy. Because
$\frac{df_p}{dt}=\frac{\partial{f_p}}{\partial{t}}+\sum\limits_{k=1}^T({V_k}
\frac{\partial{f_p}}{\partial{r_k}}+\dot{p_k}\frac{\partial{f_p}}
{\partial{p_k}})$, then from (10) for stationary case we have:
\begin{equation}
{\sum\limits_{k=1}^T({V_k}
\frac{\partial{f_p}}{\partial{r_k}}+\dot{p_k}\frac{\partial{f_p}}
{\partial{p_k}})=-f_p\sum\limits_{k=1}^T
\frac{\partial}{\partial{p_k}}F_{k}^p} \label{eqn11}
\end{equation}

The equation (11) is similarly to the equation, which described
fluctuations of the distribution function in nonequilibrium gas
[6]:
\begin{equation}
{{V_k} \frac{\partial{f_p}}{\partial{r_k}}= St{f_p}} \label{eqn12}
\end{equation}

Here, $St{f_p}$- is integral of collisions.

    Let's consider the important interrelation between descriptions of
dynamics of separate subsystems and dynamics of system as a whole.
As the expression, ${\sum\limits_{p=1}^R{\sum\limits_{k=1}^T
F_{k}^p =0}}$, is carried out, the next equation for the full
system Lagrangian, $L_R$, will have a place:
\begin{equation}
{\frac{d}{dt}\frac{\partial{L_R}}{\partial{V_k}}-
\frac{\partial{L_R}}{\partial{r_k}}=0}\label{eqn13}
\end{equation}
and the appropriate Liouville equation:
${\frac{\partial{f_R}}{\partial{t}}+{V_k}\frac{\partial{f_R}}
{\partial{r_k}}+\dot{p_k}\frac{\partial{f_R}}{\partial{p_k}}=0}$
The function, $f_R$, corresponds to the full system. The full
system is conservative. Therefore, we have: ${\sum\limits_{p=1}^R
divJ_p=0}$. This expression is equivalent to the next equality:
${\frac{d}{dt}(\sum\limits_{p=1}^{R}\ln{f_p})}=
\frac{d}{dt}(\ln{\prod\limits_{p=1}^{R}f_p})=
{(\prod\limits_{p=1}^{R}f_p)}^{-1}$
$\frac{d}{dt}(\prod\limits_{p=1}^{R}{f_p})=0$. So,
$\prod\limits_{p=1}^R{f_p}=const$. In an equilibrium state we have
$\prod\limits_{p=1}^R{f_p}=f_R$. Because the equality
$\sum\limits_{p=1}^{R}F_p=0$ is fulfilled during all time, we have
that equality, $\prod\limits_{p=1}^R{f_p}=f_R$, is a motion
integral. It is in agreement with Liouville theorem about
conservation of phase space [4]. So, only in two cases the
Liouville equation for the whole system is in agreement with the
general Liouville equation for selected subsystems: if  the
condition $\int\limits_{0}^{t}{(\sum\limits_{k=1}^T\frac{\partial}
{\partial{p_k}}F_{k}^p)}dt\rightarrow{const}$ (c) is satisfied
when $t\rightarrow\infty$, or when,
${(\sum\limits_{k=1}^T}\frac{\partial}{\partial{p_k}}F_{k}^p)$, is
a periodic function of time. The first case corresponds to the
irreversible dynamics, and the second case corresponds to
reversible dynamics. So, non-potentiality of interaction forces of
subsystems causes in an opportunity of existence of irreversible
dynamics. Let's consider, how these two types of dynamics can be
realized.

 \date{\it{Irreversible dynamics.}} \maketitle
The condition (c) occurs if the next condition is satisfied:
$\frac{\partial}{\partial{t}}F_p\leq0$, i.e.,
${F_p(0,\omega_0)}\geq{F_p(t,\omega)}=S^tF_p(0,\omega_0)$. These
conditions determine irreversibility. Here $S^t$ - a operator of
evolution or a phase stream; $\omega$ - a point of phase space.
Let's show that performance of these conditions follows from the
law of momentum conservation of colliding disks. Firstly, we
consider a simple case. Let us have $P$ pairs of colliding disks.
At that, $P$ of them are resting disks, and $P$ disks swoop on the
latter with equal velocities directed along axis $x$. Let
distances between colliding disks are equal, therefore all
collisions occur through equal interval of time, $\tau$. In this
case the total relative velocity of disks along $x$ axis before
collision would be equal to $P\Delta_k(t)$. After collisions the
inequality $P\Delta_k(t)\geq\sum\limits_{k=1}^T\Delta_k(t+\tau)$
will take place, where $\Delta_k(t)$  are projections of relative
velocity of disks on $x$ axis. Equality would take place only if
all collisions were frontal. So, as it follows from equation (2)
the force may only decrease because $F_p$ is proportional to
$\sum\limits_{k=1}^T\Delta_k$. Decreasing of $F_p$ would take
place in general case because inequality $F_p\neq0$ takes place
when the subsystem has average velocity which distinct from the
average velocities of external disks. As a result, the system
would come to stationary state and the condition for it,
$\sum\limits_{k=1}^T\Delta_k$, would be satisfied. It is
corresponds to the stationary state when $F_p=0$. From here, the
conclusion that the decrease of force, $F_p$, comes to homogeneous
distribution of system energy, is follows.

Let's consider the stability of the equilibrium state on the
example of hard-disks system, and estimate characteristic time of
decreasing force, $F_p$. For this purpose we analyze solution's
stability Eq.2 in the equilibrium point.

The evolution of the $p$-subsystem is determined by the
vector-column, $\vec{V}_T^p$, which components is speed of disks
of the $p$-subsystem: ${\vec{V}_T^p= \{{V_k^p}\}, k=1,2,3...T}$.
Some of the evolution's properties of this subsystem will be
determined by the studying of the sum of it components. Let us
designate this sum as $\Upsilon_p$. Carrying out the summation in
Eq.(2) on all disks of the $p$-subsystem, we shall obtain:
\begin{equation}
\dot{\Upsilon}_p=\sum\limits_{k=1}^T{{\Phi}_{ks}\Delta_{ks}}=
F_{P}\label{eqn14}
\end{equation}
The equation (14) describes the change of a total momentum, acting
onto the $p$-subsystem as a result of collisions. The relaxation
of a total momentum to zero is equivalent to relaxation to zero of
the force, $F_{P}$. Now let us show, if the mixing property for a
disks system is carried out, the homogeneous distribution of
impact parameters of disks have place as well. In accordance to
definition of the mixing condition, we have [7]:
$\mu(\delta)/\mu(d)=\delta/d$ where, $\mu(d)$, is a measure
corresponding to the total value of impact parameter - "$d$";
$\delta$ - is an arbitrary interval of the impact parameter and,
$\mu(\delta)$, is a corresponding measure. The fulfillment of the
mixing condition means the proportionality between the number of
collisions of disks, falling at the interval, "$\delta$", and the
length of this interval. It allows to say that distribution of the
impact parameters is homogeneous. As it is well known [7, 13], for
mixed systems the condition of depletion of correlations have
place. For the equation (14) this condition can be written down
so: $<\Phi_{ks}\Delta_{ks}>= <\Phi_{ks}><\Delta_{ks}>$ i.e.
average from two multiplied functions is equal to multiplication
of these functions average. The $\Phi_{ks}$ is dependent from
impact parameters, and $\Delta_{ks}$ is dependent from relative
velocities of colliding disks. Therefore this condition is similar
to a condition of independence of coordinates and momenta that is
widely used in the statistical physics Landau [1976]. Thus, it is
possible to execute summation in the multiplier, $\Phi_{ks}$, on
impact parameters, independent from summation of expression

$\Delta_{ks}$ on relative speeds of colliding disks. Then under
the condition of the homogeneous distribution of impact parameters
and when $T>>\infty$, we can transit from summation to
integration. So, we will have [10,11]:
$\phi=1/T\lim\limits_{T\rightarrow\infty}\
\sum\limits_{k=1}^T\varphi_{ks}=
\frac{1}{G}\int\limits_{-1}^{1}{\varphi_{ks}d(\cos\vartheta)}=-\frac{2}{3}$,
where $G=2$ is the normalization factor. Taking it into account,
we will have from eq. (14):
\begin{equation}
{\dot{\Upsilon_p}=-\frac{2}{3}\sum \limits_{k=1}^T\Delta_{ks}}.
\label{eqn15}
\end{equation}
The negative factor in the right side equation (15) means, that
the force, $F_p$, will decrease. The stability of a stationary
point $p$-subsystem can be established with the help of the motion
equation (2). Let the point, $Z_0$, be a stationary point, in
which the, $F_{P}$, effecting on $p$-subsystem, is equal to zero.
From the Lyapunov's theorem about stability follows that the
point, $Z_0$, is asymptotically stable if any deviation from it
will be attenuated. Let us expand the left and right sides of the
equation (16) into series by small parameter, $\upsilon$, of
perturbation of velocities of disks of the $p$-subsystem, near
point, $Z_0$, and keep terms to first order. The expansion of the
left side of the equation (16) gives:
$\dot{\upsilon}=\sum\limits_{k=1}^T\dot{\varepsilon}_k$, where the
summation is carried out on components of the variation,
$\upsilon$. In the expansion of the right side, the only remaining
part is,
$-\frac{2}{3}\sum\limits_{k=1}^T\varepsilon_k=-2/3\upsilon$. A
contribution into the expansion will be given by collisions of
disks of the $p$-subsystem, with disks of its complement. So, we
have: $\dot{\upsilon}=-\frac{2}{3}\upsilon.$ This equation means,
that any system deviation from an equilibrium state will decries.
Hence, the stationary point at performance of a mixing condition
is steady. Stability is provided by occurrence of returning force,
$F_p$, at a deviation of a subsystem from an equilibrium point. We
shall note, that the equation (16) also follows from the theory of
fluctuations (see Eq. [12]) where it is proved from other facts
[3].

 \date{\it{ Reversibility dynamics.}} \maketitle
According to (c) and to the general concept of reversible dynamics
the following inequalities should take place [7]:
$F_p(t+n\tau_0,\omega)=F_p(t,\omega)$,
$\Delta\Gamma_0=S^t(t+n\tau_0)\Delta\Gamma_0(t)$, where
$\Delta\Gamma_0$ is an element of volume of the phase space
occupied with the subsystem at the moment, $t$; $\tau_0$ is period
of system's return in the initial point; $n=1,2,3...;
\omega\in\Delta\Gamma_0$. This condition takes place when a vector
of the force, $F_p$ , rotates without change of its module. This
is an example, demonstrating existence of such points, there is a
system consisting of disks located on plane with a hard walls. If
at the initial moment of time this system was at phase point, in
which velocities of all disks are perpendicular to one of the
wall, and all impacts of disks are frontal; so the condition of
the reversibility would take place. I.e. in this case the system
comes back periodically to the initial point. Therefore, such
points are accepted to call as periodic or cyclic. Probability of
system's return is determined by probability to be system in these
points at the moment of preparation and by stability these points.
Really, if at the initial moment of time the system is not in one
of these points, the reversible dynamic is impossible and the
system will never come to these points. Otherwise, it might leave
this point, but that contradicts the definition of reversibility.

It is possible to assume that for system of disks at
$N\rightarrow\infty$, probability to appear in cyclic point is
very small. Then, the probability of reversible dynamic is very
small also, though it is vary from zero. Cyclic points are
determined by symmetry of boundary conditions and groups of
systems' symmetry, which can be found with help of matrix's
collision. I.e. probability of reversibility of the system depends
on geometrical characteristics of its elements and boundary
conditions. Research of these points, for example, their measures
and stability, has basic interest for process of evolution [14].
So, we can say that reversibility is possible only at the event
when at the moment of preparation of nonequilibrium system appears
in one of cyclic points. Such interpretation of reversibility
coincides with point of view of Einstein, according to which,
probabilistic description of the system is dictated by probability
to take one of the points in phase space in moment of the system
preparation. But the dynamics of the system should be determined.

I.e. the statistical description of systems is connected not with
probabilistic the nature of processes, but with opportunity of use
of such description for the analysis of dynamics of many-body
systems.

 \date{\it{ Irreversibility of rarefied systems of potentially interacting
disks.}} \maketitle Let's consider, how , the irreversibility for
rarefied systems of potentially interacting disks is follow from
the property of irreversibility for hard disks. Let us take into
account, that  pair collision approach is convenient for rarefied
system. Dynamics of potentially interacting disks is described by
the Newton's equation. In this equation the force, $F_{ks}$,
between $k$ and $s$ disks is expressed by meaning of scalar
function, $U$: $F_{ks}=-\frac{\partial{U}}{\partial{r_{ks}}}$,
where $U$ is potential energy, $r_{ks}$ is distance between disks.
It is possible to take a characteristic radius of interaction
disks, $R_{int}$ for rarefied system, which is much less than
length of free path, $l_c$, i.e. $l_c\gg{R_{int}}$. Disks can be
considered to be free when their distance  up to the nearest
neighbor disks is more than $R_{int}$. Let's consider scattering
of two disks.

In coordinate system of the centre of weights, character of
scattering is determined by the formula of [4]:
${\varphi_0}=\int^{R_{int}}_{r_{min}}\frac{\rho{dr}}
{r^2\sqrt{1-\frac{\rho_2}{r^2}-\frac{2U}{m(\Delta^0_{ks})^2}}}$.
Here, $\varphi_0$ is a scattering angle; $r$ is a distance between
scattering disks; $\rho$ is an impact parameter; $r_{min}$ is a
square root of the formula under radical; $m$ is mass of disks;
$\Delta^0_{ks}$ a velocity of swoop disk, which is equal to
relative velocity of interaction disks in laboratory co-ordinate
system. In coordinate system of the centre of weights $m=1/2$ (the
weight of a disk has been accepted to equal unit).

It is easy to show that velocity of disks, after they abandon the
area of interaction, is possible to be determined under formulas
for hard-disks, if to make replacement in them at formula of
velocity (see [4]): $\theta_{ks}=|\pi-2\varphi_0|/2,
 \cos{\theta_{ks}}=d_{ks}$.
Here, $\theta_{ks}$, $d_{ks}$, are an angles of scattering of the
disk and an impact parameter accordingly. Thus, trajectories of
disks under condition of $r\geq{R_{int}}$ are possible to be
determined with the help of formulas (2) without integration of
the Newton equation of inner area of interaction.

For rarefied system of disks the conditions,
 ${l_c}/V\gg{t_{int}}$, take place. Here, $t_{int}={R_{int}}/V$ is
characteristic time during which the disks are in the interaction
region, $V$ is characteristic velocity of disks. In this case we
can make transition, $R_{int}\rightarrow0$. Such transition
corresponds to approach of hard collisions. I.e. dynamics of the
system is described by the equation (2) rigorously. It follows
from here that dynamics of rarefied system can be studied both
with the help of the equation of Newton and the equation (2). At
$R_{int}\rightarrow0$ or $t_{int}\rightarrow0$ the calculations
results in both cases completely coincide. Hence, dynamics of
rarefied system of potentially interacting particles may be
described with help of motion equation (2). So, the rarefied
system of potentially interacting particles, as well as system of
hard colliding particles also can possess irreversible dynamics.

 \date{\it{Interrelation of the generalized field of force
with entropy.}} \maketitle If we know the generalized field of
forces, it is possible to obtain deviation of entropy for given
nonequilibrium condition of system from volume entropy of the
equilibrium state. Really, let the nonequilibrium system come to
equilibrium. Then work of the generalized field of forces would go
on increase of internal energy of the system Landau, L.D. [1976].
In this case entropy deviation, $\Delta{S}$, can be determined
with the help of following expression Somsikov, V.M. [2003]:

\begin{equation}
\Delta{S}=\sum\limits_{p=1}^R\{
\frac{T}{E_p}\sum\limits_{k=1}^T{F^p_k{dr_k}}\} \label{eqn17}
\end{equation}
where $E_p$ is kinetic energy of subsystem (all subsystems have
 ${T}$ number of the disks). I.e. with help of equations (17)
the dynamic parameter of system - generalized field of forces is
connected with thermodynamic parameter -entropy. And as the system
aspires to achieve a condition with the minimal internal work
appropriate to the minimal value of a generalized field of forces,
the equation (17) corresponds to the principle of the minimal
entropy production. It is not difficult to obtain (for example
when $R=2$) that the entropy increment for all system may be only
positive (though in some subsystems it can be negative). This
conclusion corresponds to existing results connected with entropy
change in the system at occurrence in them of some structures, for
example, turbulent structure. So, Klimontovich, Yu.L. [1] with the
help of the S-theorems, offered by him, proved decreasing of the
entropy when formations of structures have a place. The basic
difference of the formula (17) from the formula for entropy
offered by Yu.L.Klimontovich, is that the deviation entropy,
$\Delta{S}$, is determined by relative to the nonequilibrium
condition, which accepted as "back-ground"; but in the formula
(17) $\Delta{S}$ is determined by deviation of entropy an
nonequilibrium state from an equilibrium state.

\date{\it{Conclusion.}} \maketitle
So, we shown an opportunity of explanation of irreversible
dynamics on the basis of formalism of classical mechanics, if this
formalism expands by inclusion to it of the generalized Hamilton
and Liouville equations. Formally, irreversibility follows from
the right side of the generalized Liouville equation. The nature
of this equation is connected with appearing in the system of the
generalized field of forces, when it deviates from equilibrium,
and dependence its forces from particles velocities. This
dependence is caused by that that moving particles create this
field of forces. So one can submit the following explanation of
irreversibility. Let the nonequilibrium system, which velocity of
center of weights, $V_0$, is equal to zero. Because of
nonequilibrium, velocity of the center of weights some subsystems
will differ from zero. I.e. these subsystems will move relatively
each other. Then the corresponding part of kinetic energy of the
subsystems which connected with this relative subsystem moving
will be redistributed between them proportionally to their
relative velocity. This energy will go on increasing of system
entropy. As a result the relative velocity of subsystems will go
to zero and the system will come to an equilibrium state with
maximal entropy. This is a physical essence of irreversibility.

As in equilibrium state the generalized field of force is equal to
zero, so in this case, the generalized Hamilton and Liouville
equations transform into the canonical equations. And therefore in
equilibrium state Poincare's recurrence theorem (see. [13]) is
applicable. Dynamic of such systems is completely reversible. But
reversibility is possible not far from equilibrium state. So, it
is because the factors, which determining irreversibility, has the
second order of value in relation to linear disturbances.

Thus, the analysis of evolution of open nonequilibrium systems
becomes possible within the framework of the laws of classical
mechanics only on the basis of the generalized Hamilton and
Liouville equations.


\begin{thebibliography}{50}
\bibitem{Ref1}
Klimontovich, Yu.L.:  {\it{Statistical theory of the open system}}
Moscow,1995.
\bibitem{Ref2}
Lanczos, C.: {\it{The variation principles of mechanics}}, Second
edition University of Toronto Press, 1962.
\bibitem{Ref3}
Landau, L.D.: {\it{Statistical physics}} Part 1. Nauka, Moscow,
1976.
\bibitem{Ref4}
Landau, L.D. {\&} Lifshits, Ye.M.:  {\it{Mechanics}}, Nauka,
Moscow, 1973.
\bibitem{Ref5}
Lebowitz, J.L.: Boltzmann's entropy and time's arrow {\it{Physics
Today}}(1993), September, 32-38.
\bibitem{Ref6}
Lifshits, Ye.M. {\&} Pitaevsky A.P.: {\it{Phys. kinetics}}, Nauka,
Moscow, 1979.
\bibitem{Ref7}
Loskutov, F.U. {\&} Mihailov, A.S.: {\it{Introduction to
synergetic}}, Moscow, 1999.
\bibitem{Ref8}
Petrosky, T. {\&} Prigogine, I.: The Extension of classical
Dynamics for unstable Hamiltonian systems, {\it{Computers Math.
Applic.}}, V. 34. No. 2-4. (1997), 1-44.
\bibitem{Ref9} Sinai, Ya.G.: Dynamical system with
elastic reflection ergodynamic properties of scattering billiards,
{\it{Uspekhi Mat. Nauk}}, (1970), V. 25, 141-192.
\bibitem{Ref10}
Somsikov, V.M.: Non-recurrence problem in evolution of a hard-disk
system, {\it{Intern. Jour. Bifurc. And Chaos}}, 11, No 11, (2001),
2863-2866.
\bibitem{Ref11}
Somsikov, V.M.: Some approach to the Analysis of the Open
Nonequilibrium systems, {\it{AIP}} 20, (2002), 149-156.
\bibitem{Ref12}
Somsikov, V.M.: The mechanism of irreversibility in a hard-disks
system, {\it{Problems of the evolution of the open systems}}
(Almaty), V.1, (2003) 49-60.
\bibitem{Ref13}
Zaslavsky, G.M.: {\it{The stochastic of dynamics system}} (Nauka,
Moscow), 1984.
\bibitem{Ref14}
Zaslavsky, G.M. Chaotic dynamic and the origin of Statistical
laws, {\it{Physics Today}} August, Part 1, (1999), 39-45.
\end{thebibliography}
\end{document}